# Donor type semiconductor at low temperature as maser active medium

Yuri Kornyushin

*Maître Jean Brunschvig Research Unit, Chalet Shalva, Randogne, CH-3975*

In some semiconductors donor impurity atoms can attract additional electrons, forming negative donor impurity ions. Thus we have 3 energy levels for electrons: zero energy levels at the bottom of the conductivity band, negative energy levels of the bounded electrons of the negative donor impurity ions, and deeper negative energy levels of the outer electrons of the neutral donor impurity atoms. So the donor impurity atoms could serve as active centres for a maser. The maximum achievable relative population is 0.5. Typical wavelength of the generated oscillation is 0.14 mm; three level scheme could be realized at rather low temperatures, considerably lower than 6 K.

Let us consider some donor impurity atom in a donor-type semiconductor having one more outer electron than the host atom of a semiconductor (e.g., the host silicon atom). In the simplest case of a non-degenerate standard conductivity band the equation of a motion of a superfluous electron is the same as that for the electron in a hydrogen atom [1]. The bond energy at that is as follows [1]:

$$E_{bd} = (m_0 e^4/2\hbar^2)(m/m_0\varepsilon^2), \qquad (1)$$

where $m_0$ is the free electron mass, $m$ is the effective electron mass in semiconductor conductivity band, $e$ is the electron charge, and $\hbar$ is the Planck constant divided by $2\pi$.

Comparative to the bond energy in a hydrogen atom (see, e.g., [2]) the right-hand part of Eq. (1) contains additional factor $(m/m_0\varepsilon^2)$. At $m = 0.1 m_0$ and $\varepsilon = 12$ [1] this factor, $(m/m_0\varepsilon^2) = 6.944 \times 10^{-4}$.

Hydrogen atom can attract an additional electron, forming negative hydrogen ion [3]. The electron affinity of a free electron to a hydrogen atom is $E_{ah} = 0.754$ eV [3]. The bond energy in a hydrogen atom, $E_{bh} = (m_0 e^4/2\hbar^2) = 13.598$ eV [3]. Taking into account that in a donor semiconductor we have an additional factor $(m/m_0\varepsilon^2)$, we have the affinity of a conductivity band electron to a donor impurity atom $E_{ad} = 0.754(m/m_0\varepsilon^2)$ eV and $E_{bd} = 13.598(m/m_0\varepsilon^2)$ eV. At $m = 0.1 m_0$ and $\varepsilon = 12$ [1] we have $E_{ad} = 5.236 \times 10^{-4}$ eV = 6.076 K and $E_{bd} = 9.443 \times 10^{-3}$ eV.

So at temperature considerably lower than $E_{ad}$ (about 6 K here) we have donor atoms acting as active maser centres. We have three energy levels of electrons: zero energy levels at the bottom of the conductivity band, negative energy levels of electron, forming negatively charged donor ions, and deeper negative energy levels of the outer electron of the neutral donor impurity atoms. So we can pump some outer electrons of some neutral donor impurity atoms to the conductivity band. At low enough temperature these electrons will form negative impurity ions with some other neutral donor impurity atoms, thus forming highly populated levels above the ground state level, $-E_{bd}$. When high population of the upper levels is achieved, the frequency, $E_{bd} - E_{ad} = 8.92 \times 10^{-4}$ eV = $2.14 \times 10^{12}$ (1/s) = 0.14 mm, could be generated or amplified. It is rather obvious that the maximum concentration of the negative donor impurity ions, which could be achieved, is $0.5n_d$ ($n_d$ is the number of the donor impurity atoms per unit volume of a semiconductor).